# Dielectric anomalies induced by different mechanisms in $Pb(Fe_{1/2}Nb_{1/2})_{1-x}Ti_xO_3$ single crystals

**Kui Liu,[1] Xinyi Zhang,[1,*] Shiqiang Wei,[2] H. Oyanagi,[3] and Jingzhong Xiao[1,4]**

[1] *Synchrotron Radiation Research Center, Physics Department, Surface Physics Laboratory (State Key Laboratory) of Fudan University, Shanghai, 200433, China*

[2] *National Synchrotron Radiation Laboratory, University of Science and Technology of Chian, Hefei, Anhui230029, China*

[3] *Photonic Institute, National Institute of Advanced Industrial Science and Technology, 1-1-4 Umezono, Tsukuba, Ibaraki 305-8568, Japan*

[4] *International Centre for Materials Physics, Chinese Academy of Sciences, Shenyang 110016, China.*

**Abstract**

We investigated the dielectric properties of $Pb(Fe_{1/2}Nb_{1/2})_{1-x}Ti_xO_3$ single crystals below room temperature. Two dielectric anomalies were detected in sample A while only one was detected in sample B in the temperature range 90~300 K. A Debye-like relaxation with strong frequency dispersion was detected in both samples. The pre-edge XAFS suggests that this dielectric anomaly is induced by the hopping conductivity between $Fe^{2+}$ and $Fe^{3+}$. The EXAFS results give us a clear picture of the local structure of iron ions. The weak frequency dependent dielectric anomaly only observed in sample A is supposed to be due to the dipole glass behavior.

**Introduction**

Lead iron niobate $Pb(Fe_{1/2}Nb_{1/2})O_3$ (PFN) has been studied of great interest recently because of its high dielectric constant (2000~20000) and multiferroic properties which may be applied in high dielectric capacitors and multimemeory devices.[1-3] PFN belongs to the complex perovskite compounds family discovered by Smolenskii *et al.*[4] PFN is in rhombohedral phase at room temperature, and undergoes a ferroelectric-paraelectric and rhombohedral-cubic phase transition at the curie temperature ($T_C$) 385 K (Ref. 4) or 383 K (Ref. 5). PFN also exhibits long range antiferromagnetic order below the Neel temperature $T_N$~143K.[6] Y. Yang *et al* observed a magnetoelectric coupling effect in PFN single crystals. A step-like dielectric anomaly was detected near the Neel temperature (143 K).[7] $Pb(Fe_{1/2}Nb_{1/2})_{1-x}Ti_xO_3$ (PFNT) is a modified material which has been also studied extensively in recent years. PFNT undergoes a rhombohedral-tetragonal phase transition with the increase of Ti concentration.[8] Recent study reveals that the morphotropic phase boundary (MPB) is in the range $0.06 < x < 0.08$.[9] The PFNT ceramics (x=0.13) shows a diffuse phase transition at 398 K.[10] The (001)-cut PFNT (x=0.48) single crystal undergoes a tetragonal-cubic phase transition at 518 K, but no Curie temperature was detected below 570 K in the (001)-cut PFNT (x=0.06) crystal.[11,12]

Though the dielectric properties of PFNT have been studied extensively, most of the work was done above room temperature, and seldom work below room temperature was available in the literature. In this paper, we report the low temperature (90~300 K) dielectric properties of the (100)-cut PFNT single crystals. Two dielectric anomalies induced by different mechanisms were observed in the low Ti-doped sample, while only one was detected in the high Ti-doped sample. The mechanism of the dielectric anomalies were discussed based on the microstructures investigated by the Raman spectra and X ray absorption fine structure (XAFS) results.

**Experiment**

High quality PFNT single crystals (x=0.07, named sample A; x=0.48, named sample B) were fabricated without any impure phase. The detail of the single crystal growth can be found elsewhere.[13] Both samples were cut along plane parallel to (100) face. The powder X-ray diffraction patterns were measured for the phase and plane identification by a Philips X-ray diffractometer with Cu *Kα* line. The micro-Raman measurements were performed in the temperature range from 77 to 300 K by a JY HR800 spectrometer with 632.8 nm radiation. The pre-edge and extended X-ray absorption fine structure (EXAFS) data were collected at the beamline 13B of Photo Factory, National High Energy Institute of Japan at room temperature. Aluminium electrodes were evaporated on both sides of the samples to measure the dielectric properties. The dielectric measurements were carried out by a HP4284 impedance analyzer, and the probe of an AC voltage was confined to be 10 mV. Temperature dependence of dielectric spectra were recorded in the temperature range 90~300 K at various frequencies (100, 1k, 5k, 10k, 50k, 100k, 500k, 1M Hz). The temperature was defined by a temperature control system and a thermocouple. The temperature varying speed was about 1 K/min.

**Results and discussion**

Figure 1(a) shows the powder XRD patterns of both samples on the (100) plane. It can be seen that both samples are pure perovskite phase formed without any impure phase. Three peaks are observed at around 22°, 45°, 70°, corresponding to 100, 200, 300 planes, respectively. The peak-(200) of both samples is a doublet, which is usually a feature of the tetragonal phase.[9,14] As shown in the inset of Fig. 1(a), the peak-(200) of sample A and sample B are split into the peak a located at 45.1° and 44.8°, and the peak b located at 45.7° and 46.0°, respectively. The peak-(200) of rhombohedral PFN is a singlet and locates at 45.1°,[9] indicating that rhombohedral phase might exist in sample A. According to the work of S. P. Singh *et al*,[9] the sample A with x=0.07 is in the MPB region in which rhombohedral and tetragonal phases coexist. The sample B with x=0.48 is far away from the MPB region and the well separation of its peak-(200) reveals the pure tetragonal phase of sample B. Figure 1(b) shows the Raman spectra recorded at room temperature. Special attention is paid to the peaks around 275 and 800 cm$^{-1}$, which belong to the tetragonal phase $E+B_1$ mode and the rhombohedral phase $R_h$ mode, respectively.[14] The peak around 275 cm$^{-1}$ (peak-275) is the stretching mode of $Ti^{+4}$-O-$Ti^{+4}$ well match with the phonon frequency of $PbTiO_3$ in the same frequency region (281 cm$^{-1}$).[15] The much higher intensity of peak-275 of sample B further approves that the sample B is in tetragonal phase. On the other hand, the peak around 800 cm$^{-1}$ of sample A is much sharper than that of sample B, suggesting the existence of rhombohedral phase in sample A. From the XRD and Raman spectra results, we conclude that the sample A is in the MPB region with coexistence of rhombohedral and tetragonal phases, while the sample B is pure tetragonal phase formed.

Figure 2 shows the temperature dependence of dielectric spectra for both samples. Eight selected frequencies from 100 to 1M Hz are exhibited. As shown in Figs. 2(a) and 2(b), two dielectric anomalies were observed in sample A while only one was detected in sample B. The anomaly with significant frequency dependence, named the first anomaly, was detected in both samples. A step with two temperature-independent plateaus can be seen in the $\varepsilon'$-T curves, while the $\varepsilon''$-T curves shown in Figs. 2(c) and 2(d) reveal a peak in the same temperature range as the step. The dielectric constant at the low temperature plateau is only $10^2 \sim 10^3$. However, at the high temperature plateau, the dielectric constant is increased to $10^5$, exhibiting colossal dielectric constant (CDC) behavior. As the frequency is increased, the steps in $\varepsilon'$-T and peaks in $\varepsilon''$-T curves shift toward higher temperatures, showing a Debye-like relaxation. The steps move out of the temperature window when the frequency is larger than 500 kHz for sample A and 50 kHz for sample B. An Arrhenius expression can be used to describe above relaxation[16]

$$f = f_0 \exp(-E_a / k_B T_p), \qquad (1)$$

where, $f$ is the experimental frequency, $f_0$ is the preexponential factor, $T_p$ is characteristic temperature, $E_a$ is the activation energy, and $k_B$ is the Boltzmann constant. Figure 3 shows the half logarithmic plot of $f$ vs $1/T_p$. The activation energy values can be determined by fitting the experiment data by Eq. (1). The $E_a$ values of sample A and sample B are determined to be 0.155 and 0.151 eV, respectively, which are comparable to the values reported for $Ba(Fe_{1/2}Nb_{1/2})O_3$ (0.174 eV, Ref. 17), $Ba(Fe_{1/2}Ta_{1/2})O_3$ (0.176 eV, Ref. 17), $Ca(Fe_{1/2}Nb_{1/2})O_3$ (0.25 eV, Ref. 18), and $Sr(Fe_{1/2}Nb_{1/2})O_3$ (0.38 eV, Ref. 18). All the activation energy values mentioned above are comparable to that of a two site polaron hopping process of charge transfer between $Fe^{2+}$ and $Fe^{3+}$ (0.29 eV).[19] Therefore, the conductivity of our crystals is likely caused by the charge carriers hoping between $Fe^{2+}$ and $Fe^{3+}$. The existence of $Fe^{2+}$ will be further investigated by the pre-edge XAFS in the next paragraph. All the compounds mentioned above are perovskite type ferroelectrics with chemical formula $A(B_{1/2}B'_{1/2})O_3$. It is worth noticing that for the compounds with the same A site but different B site elements, for example, $Ba(Fe_{1/2}Nb_{1/2})O_3$ and $Ba(Fe_{1/2}Ta_{1/2})O_3$, close $E_a$ values are determined. However, very different $E_a$ values are determined for the compounds with different A site elements. It seems that the A-site doping can change the $E_a$ value significantly but only minor changes can be made by the B-site doping. Considering the B-site doping of our two samples, though their crystal structures are very different, close $E_a$ values are determined.

Figure 4(a) shows the background-subtracted pre-edge XAFS spectra at Fe K-edge for both samples. Three peaks (named A, B and C) guide the eyes at the first glance. The experiment data is fit perfectly by three Gaussian functions. The positions of the three peaks are determined to be 7112.8, 7114.2 and 7116.9 eV for sample A, and 7112.6, 7113.9 and 7116.8 eV for sample B, respectively. All the three peaks of sample B are slightly lower energy shifted than sample A, approximately 0.1~0.3 eV. This energy shift might be due to the different crystal structures of the two samples. The peak B is located around 7114 eV, which is close to the value reported for ferric compounds.[20] The separation between peak A and peak B corresponds to the typical separation of the order of 1.4 eV that has been reported for $Fe^{2+}$ and $Fe^{3+}$.[20] Therefore, $Fe^{2+}$ ions do exist in our crystals. The pre-edge XAFS results combined with the activation energy values suggest that the first anomaly in both samples is induced by the charge carriers hopping between $Fe^{2+}$ and $Fe^3$.

The local structures of iron ions are studied by EXAFS at Fe K-edge. Figure 4(b) shows the Fourier transform of the Fe K-edge spectra for both samples. The first coordinate shell (Fe-O) is well separated and analyzed. The structure parameters are listed in Table I and the discussions are listed below. (1) For both samples, the coordinate number is smaller than 6, suggesting the existence of oxygen vacancies. The oxygen vacancies are usually caused by oxygen deficiency during the crystal growth. Moreover, the sample A has a smaller coordinate number than the sample B, therefore, it has a higher concentration of oxygen vacancy. The higher concentration of oxygen vacancy of sample A can be also approved by the impedance spectra (not shown here) that the resistance of sample A is much smaller than that of sample B ($R_A$~60Ω; $R_B$~400Ω). (2) The Fe-O distance of sample A is only slightly larger than that of sample B, approximately 0.004 Å. (3) The Debye-Waller (DW) factor of sample A is also only slightly larger than that of sample B, indicating that the local environment of iron ions in sample A is similar to that in sample B. Based on above analyses, we can conclude that the B-site doping only has a minor impact on the local environment of iron ions. Thus, the 3$d$ electrons probably need the close energy to overcome the potential barrier to hop between $Fe^{2+}$ and $Fe^{3+}$, and the close $E_a$ values is the result. On the other hand, we suggest that the local structures of iron ions might be distorted significantly by the A-site doping. Then the potential barrier is changed considerably, resulting in the very different $E_a$ values. This suggestion is reasonable because the distance of A-Fe is much smaller than the distance of B-Fe. Therefore, the A-site doping might have a much stronger influence on the local structures of iron ions than the B-site doping.

Table I Parameters of the first coordinate shell of Fe K edge: N is coordinate number, R is the distance of Fe-O, $\sigma^2$ is Debye-Waller factor, $E_0$ is energy shift, and R is relative error.

| Sample | N | R (Å) | $\sigma^2$ (Å$^2$) | $E_0$ (eV) | R (%) |
|---|---|---|---|---|---|
| 0.07 | 5.4 | 2.011 | 0.0062 | -5.5 | 1.1 |
| 0.48 | 5.7 | 2.007 | 0.0058 | -8.2 | 2.3 |

The second dielectric anomaly (labelled by the small arrows) around 180 K was only detected in sample A, as shown in Figs. 2(a) and 2(c). The second anomaly is very different from the first one and has the following characteristics: (1) the first anomaly is significantly dependent on frequency while the second is only weakly frequency dependent; (2) the first anomaly exhibits a CDC behavior while the second just shows a small step increase in the $\varepsilon'$-T curves; (3) the second anomaly is immersed in the first one when the frequency is smaller than 10 kHz. It is well known that ferroelectrics near the MPB region usually have a mixed phase structure, which may induce a dipole glass behavior.[21] As discussed previously, the sample A with coexistence of rhombohedral and tetragonal phases is near the MPB region, while the sample B with pure tetragonal phase is far away from the MPB region. Therefore, based on above analyse, we suggest that the second anomaly is probably induced by the dipole glass behavior. Figure 5 shows the temperature dependence of loss tangent (tanδ) of sample A. A peak corresponding to the second anomaly can be observed around $T_m$~180 K. The peak shifts toward higher temperature with the increase of frequency. The frequency dependence of $T_m$ can be described by the Vogel-Fulcher law[22]

$$f = f_0 \exp\left(\frac{-E_a}{k_B(T_m - T_f)}\right), \qquad (2)$$

where, $f$ is the experimental frequency, $f_0$ is the preexponential factor, $E_a$ is the activation energy, $k_B$ is the Bolzmann constant, and $T_f$ is the static freezing temperature. The inset of Fig. 5 shows the Vogel-Fulcher fit of $T_m$ for sample A. From the nonlinear fitting of our experimental data, $E_a$, $f_0$, and $T_f$ are determined to be 1.25 x10$^{-2}$ eV, 2.6x10$^{10}$ Hz, and 171 K, respectively. The freezing temperature matches well with the observed peak in tanδ.

Recently, the similar anomaly was observed in $Pb(Zr_{1-x}Ti_x)O_3$, and it was suggested to be due to oxygen vacancies and related effects.[23] However, it is not the case in sample A, because the second anomaly can not be observed in sample B which also has oxygen vacancies. Since PFNT exhibits multiferroic properties, dielectric anomaly may be induced by magnetoelectric coupling effect near the Neel temperature. The temperature dependence of DC and AC magnetic susceptibility was recorded from 5K to 300K, but no ferromagnetic phase transition was found around 180 K. Moreover, the temperature dependence of Raman spectra reveal that there is no structure phase transition in sample A in the temperature range 77~300 K. Therefore, the possibility of the second anomaly induced by oxygen vacancies and related effects, magnetoelectric coupling effect or structure phase transition can be ruled out.

**Conclusions**

In summary, the microstructures and dielectric properties of PFNT single crystals were investigated. Two dielectric anomalies were detected in sample A while only one was detected in sample B in the temperature range 90~300 K. The first anomaly with strong frequency dispersion was detected in both samples. It can be attributed to the charge carriers hopping between $Fe^{2+}$ and $Fe^{3+}$. The pre-edge XAFS and EXAFS did a good job to give us a clear picture of the localized state hopping conductivity. The similar local environment of iron ions is the origin of the close activation energy of the two samples. The second anomaly observed in sample A is weakly dependent on frequency, and is supposed to be induced by the dipole glass behavior.

**Acknowledgements**

This work is supported by the National Natural Science Foundation of China (10401004). Professor F. Lu, doctor Q. J. Cai and L. X. Sun are gratefully acknowledged for the dielectric and Raman spectra measurement. The authors also thank Prof. H. Luo of Shanghai Institute of Ceramics (CAS) for his providing us part of samples.

**Figure captions**

Figure 1 (color on line) (a) Powder XRD patterns from (100) plane; (b) Raman spectra of sample A and sample B at room temperature.

Figure 2 Temperature dependence of dielectric constant and dielectric loss (Eight frequencies are exhibited: 100, 1k, 5k, 10k, 50k, 100k, 500k 1MHz). The direction of the big arrow stands for frequency increasing.

Figure 3 Arrhenius plot of the frequency dependence of the peak temperature in $\varepsilon''$-T curves.

Figure 4 (a) Pre-edge absorption spectra of Fe K edge, symbols are experiment data, and lines are fitting results by Gaussian function; (b) Fourier transform of Fe K edge (first coordinate shell), symbols are experiment data, and lines are fitting results.

Figure 5 (color on line) Temperature dependence of the loss tangent of sample A at various frequencies: 50, 100, 500 kHz and 1 MHz. The inset is the Vogel-Fulcher fit of the frequency dependence of $T_m$.

Figure 1

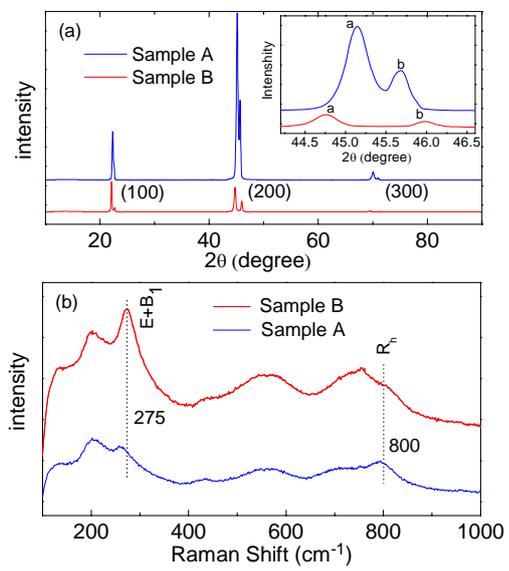

Figure 2

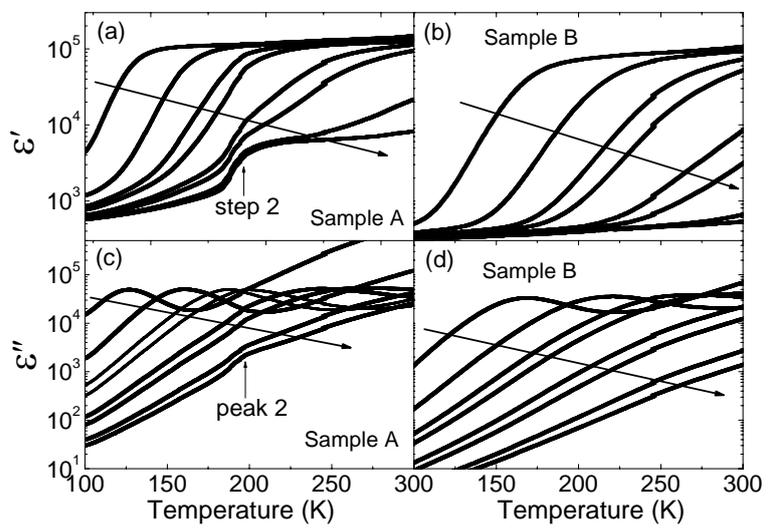

Figure 3

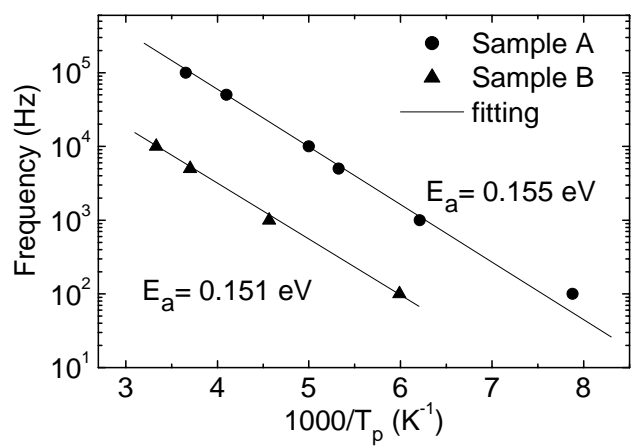

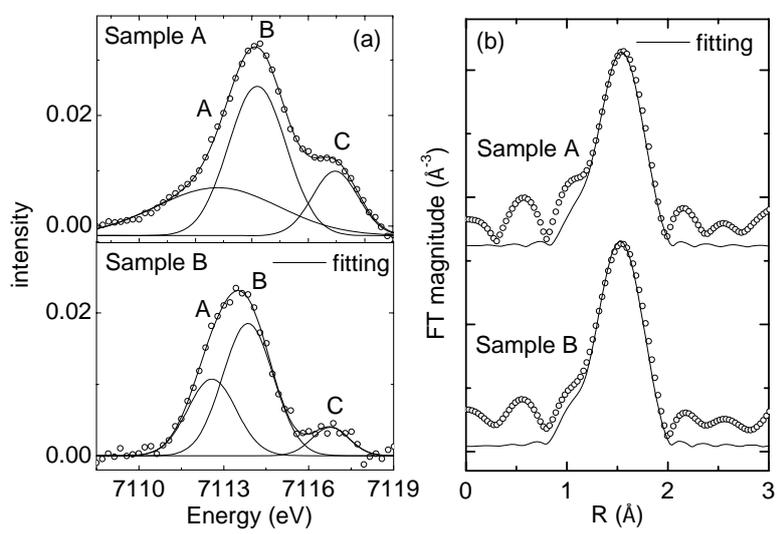

Figgure 5

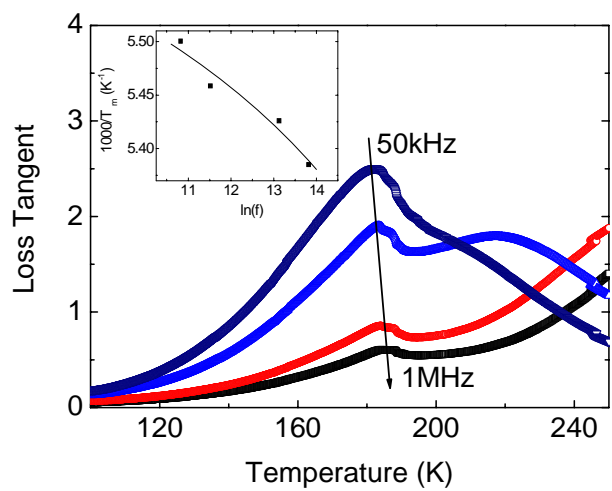